\documentstyle[psfig]{mn}

\def\ga{\mathrel{\hbox{\rlap{\hbox{\lower4pt\hbox{$\sim$}}}\hbox{$>$}}}}

\title[High-speed photometry of the Recurrent Nova IM Normae]
{High-speed photometry of the Recurrent Nova IM Normae}
\author[Patrick A. Woudt and Brian Warner]
       {Patrick A. Woudt\thanks{E-mail: pwoudt@circinus.ast.uct.ac.za} 
        and Brian Warner\thanks{E-mail: warner@physci.uct.ac.za}\\
        Department of Astronomy, University of Cape Town, Private Bag,
        Rondebosch 7700, South Africa}
\date{}

\begin{document}

\maketitle

\begin{abstract}
      The recurrent nova IM Nor is found to have an orbital period of 2.462 h, 
shown by periodic dips in brightness. This is only the second recurrent nova 
known to have such a short period. We interpret the light curve as largely 
produced by a reflection effect from the heated face of the secondary star, 
probably with addition of a partial eclipse of the accretion disc.
\end{abstract}


\section{Introduction}

    The light curve of Nova Normae 1920 was constructed from archived 
plates by Elliot \& Liller (1972), who noted that it could be grouped with 
slow novae such as DQ Her and T Pyx. A poorly determined original 
position resulted in no certain identification of the remnant  (Duerbeck 
1987), which was designated IM Nor. A second eruption in January 2002 
(Liller 2002) showed that IM Nor is in fact a recurrent nova and the 
improved accuracy of position enabled the pre-nova to be identified as a star 
with variations in brightness around V $\sim$ 18 (Kato et al.~2002). The eruption 
light curve derived by Kato et al.~led them to conclude that although it is 
indeed like T Pyx, a better comparison may be made with CI Aql, which is 
another recurrent nova. The spectra obtained during outburst show that all of 
these stars resemble each other in having strong Fe II emission and late 
appearance of forbidden lines (Kato et al.~2002; Duerbeck et al.~2003). Kato 
et al.~proposed that CI Aql and IM Nor form a distinct subclass with massive 
ejecta and long recurrence intervals.
    
   At maximum light IM Nor reached V $\sim$ 7.8 and decayed as a fast nova 
with $t_2 \sim 20$ d. The eruption amplitude of $\sim$ 10 mag agrees with that expected 
for a fast nova of low orbital inclination (see Fig. 5.4 of Warner 1995). 
However, CI Aql also showed such an initial steep fall, but this was 
succeeded by a long-lived phase of slow decline (Matsumoto et al.~2001), 
uncharacteristic of a fast nova. IM Nor is behaving similarly, so in neither 
case can the usual Amplitude/Rate of Decline/Inclination relationship be 
trusted. CI Aql is an eclipsing system with a period of 14.8 h (Mennickent \& 
Honeycutt 1995).

    IM Nor is currently at V $\sim$ 16.5 and is thus still far from its quiescent 
magnitude. We have observed it photometrically with good time resolution 
in order to look for any orbital modulation.

\begin{table*}
 \centering
  \caption{Observing log.}
  \begin{tabular}{@{}llrrrrrcc@{}}
 Object       & Type         & Run No.  & Date of obs.          & HJD of first obs. & Length    & $t_{in}$ & Tel. &  V \\
              &              &          & (start of night)      &  (+2452000.0)     & (h)       &     (s)   &      & (mag) \\[10pt]
{\bf IM Nor}  & RN           & S6807    & 2003 Feb 25 &  696.46942  &   2.93      &       20  &  40-in & 16.5\\
              &              & S6811    & 2003 Feb 26 &  697.42046  &   1.56      &       30  &  40-in & 16.6\\
              &              & S6813    & 2003 Feb 26 &  697.54623  &   2.35      &       30  &  40-in & 16.5\\
              &              & S6816    & 2003 Feb 27 &  698.41376  &   5.14      &       30  &  40-in & 16.5\\
              &              & S6824    & 2003 Mar 02 &  701.55380  &   2.25      &       30  &  40-in & 16.6\\[5pt]
\end{tabular}
{\footnotesize 
\newline 
Notes: RN = Recurrent Nova; $t_{in}$ is the integration time.\hfill}
\label{tab1}
\end{table*}

\section{Photometric observations}

               Our observations were made in white light using the UCT CCD 
Photometer (O'Donoghue 1995) attached to the 40-in reflector at the 
Sutherland site of the South African Astronomical Observatory. The 
observing log is given in Table 1 and the reduced light curves are shown in 
Figure 1.

\begin{figure*}
\centerline{\hbox{\psfig{figure=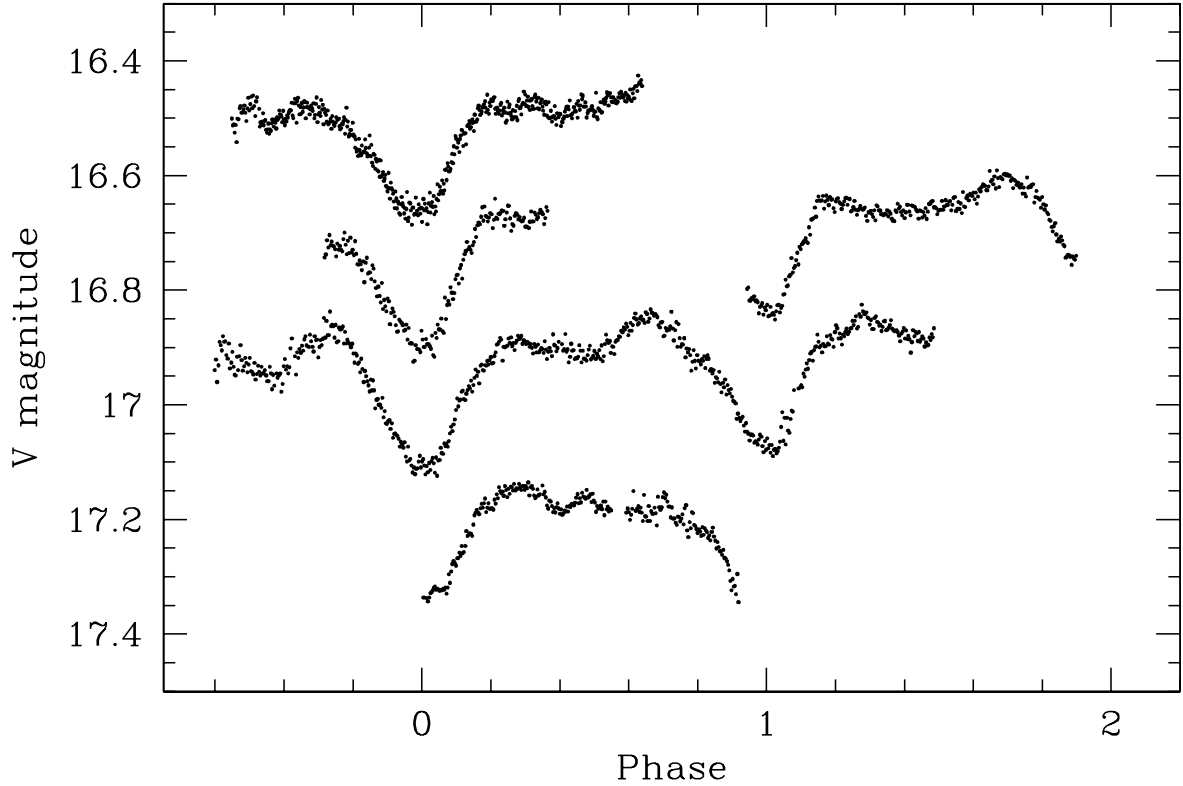,width=16.0cm}}}
  \caption{The light curves of IM Nor, obtained in February 2003. The upper light curve is
at the correct brightness, the others have been displaced vertically downwards by 0.2 mag consecutively
for display purposes only.}
 \label{lcimnor}
\end{figure*}

     IM Nor has a clear periodic brightness modulation with a range of $\sim$ 0.3. 
The mean brightness of the system varies by at least 0.1 mag from night to 
night, and can vary by that amount over a few hours. The dominant feature 
is an eclipse-like recurrent dip, with a depth of 0.2 mag, from which a period 
of 2.462 h is derived. The ephemeris for this dip is

\begin{equation}
{\rm HJD_{min}} = 245\,2696.526 + 0.1026 (\pm 1) \, {\rm E}.
\label{eq1}
\end{equation}

    Interpreted as an orbital period, the modulation at 2.462 h places IM Nor 
towards the low end of the range of orbital periods for novae and squarely in 
the centre of the `period gap' that exists for dwarf novae (but probably not 
for novae: Warner (2002)). Of the nine known recurrent novae only T Pyx 
($P_{orb}$ = 1.83 h) and IM Nor have orbital periods at the lower end.

   In the Fourier transforms of individual nights, and the whole data set, no 
periods other than the 2.462 h modulation and its harmonics were found.

\begin{figure}
\centerline{\hbox{\psfig{figure=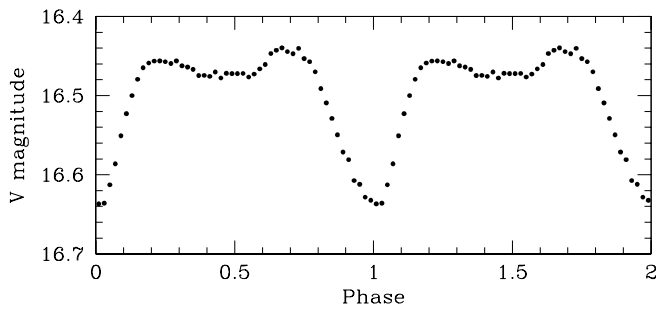,width=8.8cm}}}
  \caption{The average light curve of IM Nor in February 2003.}
 \label{avnor}
\end{figure}

\section{Discussion}   

    The width of the main brightness dip is too large for it to be the eclipse of 
an accretion disc (or of the primary, even though possibly bloated by the 
effects of the recent eruption). To assist in interpretation we examine the 
light curve of CI Aql during its decline from eruption. Figure 4 of 
Matsumoto et al.~(2001) shows the folded light curve during the decline. This 
light curve is described in that paper as a `primary eclipse with very broad 
wing extending to $\sim$ 0.7 orbital phase duration'. The whole of this feature 
cannot of course be an eclipse -- it is impossibly wide. We suggest that the 
light curve is that resulting from a moderate reflection effect from the 
secondary -- irradiated by the very hot primary left after eruption. This 
produces a single hump per orbital period, and the minimum of the reflection 
effect may be pulled even lower by partial eclipse of the accretion disc by 
the secondary at that phase.

    During quiescence, in the absence of the reflection effect, CI Aql has a 
primary eclipse of normal width that is in phase with the minimum seen 
during decline from eruption (Lederle \& Kimeswenger 2003).

    Using the same model for IM Nor we see that a single humped orbital 
modulation, again probably enhanced in depth at the minimum phase by 
partial eclipse of the accretion disc, can account for the major modulation 
around orbit. But in IM Nor there is the additional occurrence of a dip 
between maxima, with its minimum close to phase 0.5, which creates the 
rise of brightness seen before the onset of the major dip. This asymmetric 
feature may be a secondary eclipse caused by partial obscuration of the 
brightly lit hemisphere of the secondary by the optically thick accretion disc 
and the accretion stream. We have previously drawn attention to the 
possibility of such an effect in systems with strongly irradiated secondaries, 
and that such an effect might be weakly present in the nova remnant DD Cir 
(Woudt \& Warner 2003).

    The short orbital period for IM Nor requires another model to be explored. 
At this period, and with the high rate of mass transfer expected for a recent 
nova, which is irradiating its secondary, the accretion disc may be excited 
into an elliptical shape and the observed recurrent humps could be due to 
superhumps (Warner 1995). The observed amplitude of 0.3 mag is 
compatible with a superhump, but at that amplitude superhumps seen in 
dwarf nova outbursts are usually triangular in profile. The superhumps seen 
in other recent novae of short period, namely V1794 Cyg (Retter, Leibowitz 
\& Ofek 1997), DD Cir (Woudt \& Warner 2003), V4633 Sgr (Lipkin et al.~2001), 
have much lower amplitudes ($\sim$ 0.05 mag).  IM Nor has at least 1.5 
mag of decaying nova luminosity added to its quiescent brightness, which 
should result in dilution of any superhump modulation. We note that the 
basic profile of the modulation remains roughly symmetrical about 
minimum light: i.e., there is no sign of a superhump drifting relative to an 
orbital eclipse as seen in some nova remnants (e.g. RR Cha: Woudt \& 
Warner 2002). We conclude that the modulation seen in IM Nor is unlikely 
to be due to superhumps.

\section*{Acknowledgments}
PAW is supported by funds made available from the National Research
Foundation and by strategic funds made available to BW from the
University of Cape Town. BW's research is supported by the University.

\end{document}